\documentclass[%
reprint,
superscriptaddress,
 amsmath,amssymb,
prb,
]{revtex4-2}
\usepackage{amssymb}
\usepackage{amsmath}
\usepackage{bbm}
\usepackage{bm}
\usepackage{epsfig}
\usepackage{color}
\usepackage{multirow}
\usepackage{graphicx}
\usepackage{booktabs}
\usepackage{dsfont}
\usepackage{tikz}
\usepackage{appendix}
\usepackage{makecell}
\usepackage{subfigure}
\usepackage{braket}
\usepackage{xcolor}
\usepackage[colorlinks,linkcolor=blue,anchorcolor=blue,citecolor=blue,urlcolor=blue]{hyperref}
\begin{document}

\title{Intrinsic Mirror Symmetry and Robustness of Optimal Nonlocal Operators in One-Dimensional Quantum Spin Chains}

\author{Jia Bao}
\affiliation{Department of Physics, Wuhan University of Technology, Wuhan 430070, China}

\author{Bin Guo}
\email{binguo@whut.edu.cn}
\affiliation{Department of Physics, Wuhan University of Technology, Wuhan 430070, China}

\author{Shu Qu}
\affiliation{School of Electrical and Electronic Engineering, Wuhan Polytechnic University, Wuhan 430023, China}

\author{Fanqin Xu}
\affiliation{School of Electrical and Electronic Engineering, Wuhan Polytechnic University, Wuhan 430023, China}

\author{Xueyi Lei}
\affiliation{School of Electrical and Electronic Engineering, Wuhan Polytechnic University, Wuhan 430023, China}

\author{Zhaoyu Sun}
\affiliation{School of Electrical and Electronic Engineering, Wuhan Polytechnic University, Wuhan 430023, China}

\date{\today}

\begin{abstract}
Multipartite nonlocality has been extensively investigated within one-dimensional quantum lattices. Previous research has primarily focused on the nonlocality measure $S$, which quantifies the violation of Bell-type inequalities. However, the optimal nonlocal operators, which are related to specific experimental settings required to achieve the violation, often remain elusive. In this work, we employ a string-like nonlocal operator $\hat{S}_N$, characterized by a core single-site operator $\hat{p}$, to investigate the optimal  measurement setting in translationally invariant quantum chains. By analyzing the infinite-size transverse-field Ising, Cluster-Ising, and extended Ising models, we uncover two general results. First, for typical ground states, we find that the optimal single-site operator $\hat{p}$ possesses an intrinsic mirror symmetry. Second, the optimal nonlocal operator $\hat{S}(\hat{p})$ exhibits remarkable robustness: for a specific model, as the Hamiltonian parameter changes, the structure of $\hat{p}$ remains stable and persists across distinct quantum phases. These findings not only redefine the numerical optimization paradigm for multipartite nonlocality, but also significantly simplify the experimental requirements by identifying fixed measurement bases. This structural stability provides practical guidance for implementing macroscopic Bell tests in large-scale quantum simulators, making it highly compatible with modern efficient measurement protocols.
\end{abstract}

\maketitle

\section{Introduction \label{sec:intro}}

Quantum nonlocality is a cornerstone of quantum mechanics, representing a fundamental departure from classical local-realistic descriptions of the world~\cite{Brunner2014, Popescu2014, Gebhart2021, PozasKerstjens2022, Banaszek1999, Forster2009, Luo2011, Gallego2012}. In recent years, the study of nonlocality in many-body systems has gained significant momentum, fueled by the rapid development of programmable quantum simulators and high-fidelity experimental platforms~\cite{Aolita2012, Bowles2016, ContrerasTejada2021, Rowe2001, Storz2023, Drummond1983, Walther2005a}.

Multipartite nonlocality is typically identified through the violation of Bell-type inequalities. For $N$-partite spin-$1/2$ systems, the Mermin-Klyshko-Svetlichny (MKS) operators play a central role in detecting these correlations~\cite{Scarani2001, Collins2002a}. However, the inherent complexity of optimizing multivariable Bell inequalities has necessitated the use of sophisticated numerical methods. Early investigations into the nonlocality measure $S$ in quantum spin chains were restricted to small system sizes~\cite{Campbell2010}. The subsequent integration of tensor network techniques enabled the exploration of scaling behavior in much larger systems~\cite{Sun2015}. A significant breakthrough was the formulation of a nonlocality transfer-matrix theory~\cite{Xu2024, Sun2022}, where the nonlocality spectrum $\{|\lambda_i|\}$, defined by the eigenvalues of the transfer matrix, serves as a powerful tool for analyzing the thermodynamic limit of the Bell violation.

Despite these theoretical advances, a significant gap remains between numerical results and experimental realization. While the nonlocality measure $S$ and its spectrum $\{|\lambda_i|\}$ quantify the magnitude of Bell violations, they provide little insight into the configuration of the optimal nonlocal operators (NLOs) that achieve these maximums~\cite{Sun2014b, Sun2015, Sun2023a, Bao2020a, Cheng2020, Liu2021, Sun2021b, Sun2021c, Sun2021}. This lack of clarity poses a formidable challenge for experimental realization. Although Bell tests have been performed for up to $N = 14$ qubits~\cite{Lanyon2014}, scaling to larger systems is hindered by the difficulty of identifying the specific measurement settings required to maximize violations. Recent work has begun to address this in spin-1 systems~\cite{Aloy2026}, yet for the ground states of  spin-$1/2$ quantum chains, the optimal NLO configurations have remained largely elusive.

In this work, we bridge this gap by explicitly characterizing the optimal NLOs for the ground states of several prototypical one-dimensional spin-$\frac{1}{2}$ chains: the transverse-field Ising model~\cite{Pfeuty1970, Liu2021}, the Cluster-Ising model~\cite{Giampaolo2014, Qu2025}, and the extended Ising model~\cite{Liu2025d}. Utilizing an improved transfer-matrix framework~\cite{Xu2025a}, we represent the NLO $\hat{S}_N$ in a string-like form governed by a core single-site operator $\hat{p}$, which allows us to decouple the local measurement settings from the global entanglement structure. Our analysis reveals two striking features of the optimal operator $\hat{p}$. First, we identify a universal mirror symmetry with respect to a principal axis, allowing us to represent $\hat{p}$ as structured, intuitive forms. Second, we find that in each model, the optimal $\hat{p}$ operator exhibits remarkable robustness, remaining stable even as the ground states undergo quantum phase transitions driven by the controlling parameter in the Hamiltonian. These findings fundamentally redefine the numerical optimization paradigm for multipartite nonlocality, and provide a practical, fixed-basis blueprint for implementing scalable Bell tests in experiments, alleviating the need for extensive real-time measurement optimizations.

The remainder of this paper is organized as follows: Section~\ref{sec:concepts_formulas} reviews the essential concepts of multipartite nonlocality, including its spectrum, and proposes numerical and analytical methods to analyze the optimal $\hat{p}$ operator. Section~\ref{sec:results} presents the numerical results for the optimal $\hat{p}$ operators across the three aforementioned Ising-type models. Finally, Sec.~\ref{sec:conclusion} summarize and discusses our findings.

\section{Theory and Methods}\label{sec:concepts_formulas}

\begin{figure}
\centering
\includegraphics[width=0.45\textwidth,keepaspectratio]{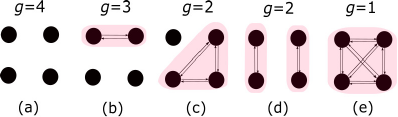}
\caption{Multipartite correlation hierarchy in a four-site system. Colored shadows denote $g$ distinct groups, within which correlations are localized. A smaller value of the group number $g$ denotes a higher  hierarchy of multipartite correlations. From left to right: a fully separable state ($g=4$, no nonlocal correlations); states with some hierarchy of multipartite correlations ($g=3$ and $g=2$); and genuine multipartite nonlocality ($g=1$, all sites sharing a single correlated entity). }
\label{Fig1}
\end{figure}

We begin by reviewing the fundamental concepts of multipartite nonlocality in Sec.~\ref{sec:group_number}, followed by an introduction to the nonlocality spectrum and the associated NLOs within the transfer-matrix formalism in Sec.~\ref{sec:transfer_matrix}. Then, we propose an alternative numerical approach to characterize the behavior of nonlocality spectrum in Sec.~\ref{sec:approximation}. Finally, we outline the analytical framework used to determine the optimal operator $\hat{p}$ in Sec.~\ref{sec:pattern_analysis}.

\subsection{Group number and multipartite nonlocality}\label{sec:group_number}

To quantitatively characterize multipartite quantum correlations, integer indicators provide an intuitive and rigorous metric~\cite{Bancal2009}. Figure~\ref{Fig1} schematically illustrates distinct patterns of multipartite correlations for a system of $N=4$ sites. The study of multipartite nonlocality in $N$-qubit systems extensively utilizes MKS operators and Bell-type inequalities to establish bounds on the group number $g$. For each site $j$ ($1 \le j \le N$), we define two local measurement operators: 
\begin{equation}\label{Eq1}
\hat{m}_{j}=\boldsymbol{a}_{j}\cdot \boldsymbol{\sigma}, \ \hat{m}'_{j}=\boldsymbol{a}'_{j}\cdot \boldsymbol{\sigma}, 
\end{equation}
which $\boldsymbol{a}_j$ and $\boldsymbol{a}'_{j}$ are unit vectors in $\mathbb{R}^3$, and $\boldsymbol{\sigma}=[\hat{\sigma}_{x},\hat{\sigma}_{y},\hat{\sigma}_{z}]$ denotes the vector of Pauli matrices. The MKS operators are defined recursively, starting from $\hat{M}_{1}=\hat{m}_{1}$ and $\hat{M}'_{1}=\hat{m}'_{1}$~\cite{Bancal2009,Collins2002,Mermin1990}:
\begin{align}\label{Eq2}
   \hat{M}_{i}=\frac{1}{2}\hat{M}_{i-1}\otimes(\hat{m}_{i}+\hat{m}'_{i})+\frac{1}{2}\hat{M}'_{i-1}\otimes(\hat{m}_{i}-\hat{m}'_{i}).
\end{align}

For an arbitrary $N$-qubit system, if the correlations admit a description by a model with at most $g$ groups, the expectation values of these operators satisfy the following Bell-type inequalities~\cite{Bancal2009}:
\begin{align}\label{Eq3}
   S = \begin{cases}
     \max_{\{\boldsymbol{a}\}}\langle \hat{M}_{N} \rangle  \le 2^{\frac{N-g}{2}} , &  \text{for } N-g \text{ even},\\
     \max_{\{\boldsymbol{a}\}}\langle \hat{S}_{N} \rangle  \le 2^{\frac{N-g}{2}} , & \text{for } N-g \text{ odd}.
\end{cases}
\end{align}

Here $S$ denotes the expectation value maximized over all measurement settings $\{\boldsymbol{a}\}$; in the even case it is associated with $\hat{M}_{N}$, and in the odd case with $\hat{S}_{N} = \frac{1}{\sqrt{2}}(\hat{M}_{N} + \hat{M}'_{N})$. For brevity, we employ the unified symbol $S$, with the understanding that the relevant operator is determined by the parity of $N-g$.

The notation $\{\boldsymbol{a}\} = \{\boldsymbol{a}_1, \boldsymbol{a}'_1, \ldots, \boldsymbol{a}_N, \boldsymbol{a}'_N\}$ represents the complete set of $2N$ unit vectors specifying the measurement settings. A violation of Eq.~(\ref{Eq3}) for a given $g$ signifies that the system's correlations cannot be partitioned into $g$ distinct groups, implying that they require a description by at most $g-1$ groups. Consequently, the true group number of a state is determined by finding the smallest $g$ for which the inequality holds, or equivalently, the largest $g$ for which it is violated. We note that while violation constitutes a sufficient condition for multipartite nonlocality, it is not strictly necessary.

\begin{figure}
\centering
\includegraphics[width=0.45\textwidth,keepaspectratio]{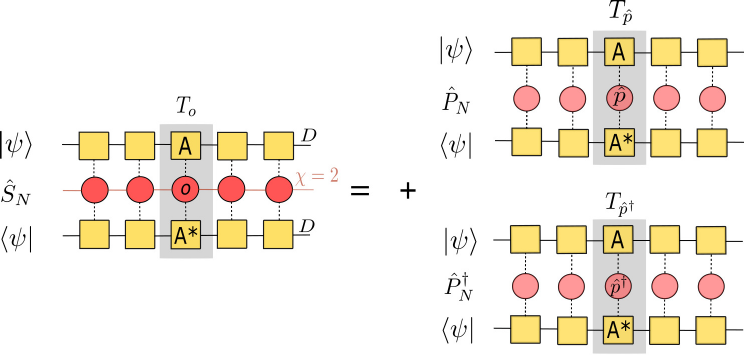}
\caption{A transfer matrix theory for the nonlocality measure $\langle \psi\vert \hat{S}_N \vert\psi\rangle$ in quantum chains with single-site unit cells (i.e., $u=1$)~\cite{Xu2025a}.      The translation-invariant ground state $|\psi\rangle$ is represented as a Matrix Product State (MPS) composed of repeating local tensors $A$. The MKS operator $\hat{S}_N$ is decomposed into the sum of two string-like operators, $\hat{P}_N$ and $\hat{P}^{\dagger}_N$, both uniquely defined by the single-site operator $\hat{p}$. This operator $\hat{p}$ serves as the fundamental constituent of the nonlocality transfer matrices ($T_{\hat{p}}$ and $T_{\hat{p}^{\dagger}}$) and dictates the scaling properties of the measure. The characterization of $\hat{p}$, which depends upon two measure directions $\boldsymbol{a}$ and $\boldsymbol{a}'$ [see Eq.~(\ref{Eq12})], is the primary focus of this study.}
\label{Fig2}
\end{figure}

\subsection{Transfer-matrix theory for the nonlocality spectrum and nonlocal operators}\label{sec:transfer_matrix}

For the ground states $\ket{\psi}$ of one-dimensional, infinite-size, translation-invariant quantum chains, the large-$N$ behavior of nonlocality can be effectively analyzed using transfer-matrix theory. The nonlocality measure $S$ is defined as the normalized expectation value
\begin{align}\label{Eq4}
    S = \frac{\bra{\psi}\hat{S}_N\ket{\psi}}{\braket{\psi|\psi}}.
\end{align}

As detailed in Ref.~\cite{Xu2025a}, this approach leverages tensor networks. As illustrated in Fig.~\ref{Fig2}, we employ the Matrix Product State (MPS)~\cite{Cirac2021, Verstraete2008, Orus2014} representation for the ground state $\ket{\psi}$ and the Matrix Product Operator (MPO)~\cite{Pirvu2010} representation for the MKS operator $\hat{S}_N$, which is expressed as
\begin{align}\label{Eq5}
   \hat{S}_N = o \cdot o o \cdots o o\cdot o,
\end{align}
where $o(\boldsymbol{a}, \boldsymbol{a'})$ is the core tensor given by
\begin{align}\label{Eq6}
   o(\boldsymbol{a}, \boldsymbol{a'}) = \frac{1}{2} \begin{pmatrix} \hat{m} + \hat{m}' & \hat{m}' - \hat{m} \\ \hat{m} - \hat{m}' & \hat{m} + \hat{m}' \end{pmatrix},
\end{align}
with $\hat{m}$ and $\hat{m}'$ (refer as Eq. (\ref{Eq1})) being site-independent by translation invariance. Although Fig.~\ref{Fig2} depicts a single-site unit cell ($u=1$), this representation generalizes straightforwardly to larger unit cells (e.g., $u=2$ with the pattern $-A-B-$).

By exploiting the translation invariance of the ground state, the expectation value $\langle\psi|\hat{S}_N|\psi\rangle$ can be efficiently evaluated through the contraction of transfer matrices $T_o$, formed by contracting the core tensor $o$ with the conjugated state tensor. This approach, rooted in the MPS formalism~\cite{Cirac2021, Verstraete2008, Orus2014}, allows us to access the thermodynamic limit without exponential overhead. Specifically, the expectation value is given by
\begin{align}\label{Eq7}
   \langle \psi | \hat{S}_N | \psi \rangle \approx \text{Tr}(T_o^N) = \sum_i d_i |\lambda_i|^N,
\end{align}
where $\{|\lambda_i|\}$ are the eigenvalues of $T_o$, sorted in descending order of magnitude. We refer to these as the \textit{nonlocality spectrum}. For large $N$, the sum is dominated by the leading eigenvalue $|\lambda_1|$:
\begin{equation}\label{Eq8}
\langle \psi | \hat{S}_N | \psi \rangle \sim d_1 |\lambda_1|^N.
\end{equation}

Combining this with Eq.~(\ref{Eq3}), the group number $g$ is bounded by
\begin{align}\label{Eq9}
   g \lesssim N(1 - 2 \log_2 |\lambda_1|).
\end{align}

This relation emphasizes that $|\lambda_1|$ directly controls the extent of multipartite nonlocality. The quantum mechanical upper bound is $|\lambda_1| \le \sqrt{2}$~\cite{Xu2025a}. A value $|\lambda_1| > 1$ implies $g < N$, signaling the presence of non-trivial multipartite nonlocality.

To reveal the hidden structure of the NLOs, we perform the decomposition $o=V\Omega V^{-1}$ in Eq.~(\ref{Eq5}). This transforms the MPO into the form~\cite{Xu2025a}
\begin{align}\label{Eq10}
\hat{S}_N &= o \cdot o o \cdots o o\cdot o \nonumber\\
&= o \cdot V\Omega V^{-1}  \cdot V\Omega V^{-1} \dots V\Omega V^{-1} \cdots o\nonumber\\
&= \Omega \cdot \Omega \Omega \cdots \Omega \Omega \cdot\Omega,
\end{align}
where $\Omega$ is the diagonal matrix of eigenvalues of the $2 \times 2$ matrix $o$:
\begin{align}\label{Eq11}
\Omega = \begin{pmatrix}
    \hat{p} & 0 \\
    0 & \hat{p}^\dagger
\end{pmatrix} 
\end{align}

Specifically, the single-site operators $\hat{p}$, which depends upon two measurement directions $\boldsymbol{a}$ and $\boldsymbol{a}'$, is expressed as:
\begin{align}\label{Eq12}
   \hat{p} = \frac{1-i}{2}\hat{m} + \frac{1+i}{2}\hat{m}' = \frac{1-i}{2}\boldsymbol{a}\cdot\boldsymbol{\sigma} + \frac{1+i}{2}\boldsymbol{a}'\cdot\boldsymbol{\sigma}.
\end{align}

With this diagonalized MPO form of $\hat{S}_N$, we obtain
\begin{equation}\label{Eq13}
\langle\psi|\hat{S}_N|\psi\rangle = T_{\Omega} \cdot T_{\Omega}T_{\Omega} \cdots T_{\Omega}T_{\Omega} \cdot T_{\Omega}.
\end{equation}

Thus, $\langle\psi|\hat{S}_N|\psi\rangle$ is determined by the eigenvalue spectrum of the new transfer matrix $T_{\Omega}$. Owing to the diagonal structure of $\Omega$, the matrix $T_{\Omega}$ also acquires a block-diagonal form:
\begin{equation}\label{Eq14}
T_{\Omega} =
\begin{pmatrix}
T_{\hat{p}} & 0 \\
0 & T_{\hat{p}^\dagger}
\end{pmatrix},
\end{equation}
where $T_{\hat{p}}$ and $T_{\hat{p}^\dagger}$ are related by $T_{\hat{p}} = X T_{\hat{p}^\dagger}^* X^{-1}$, with $X$ denoting an elementary row (column) permutation of $T^{*}_{\hat{p}^\dagger}$.

To provide an intuitive physical picture of the optimal NLOs, we note that the chain product of diagonal matrices $\Omega$ takes the form
\begin{align}\label{Eq15}
\Omega\Omega\cdots\Omega &=
\begin{pmatrix}
\hat{p} & 0 \\
0 & \hat{p}^\dagger
\end{pmatrix}
\begin{pmatrix}
\hat{p} & 0 \\
0 & \hat{p}^\dagger
\end{pmatrix}
\cdots
\begin{pmatrix}
\hat{p} & 0 \\
0 & \hat{p}^\dagger
\end{pmatrix} \nonumber\\
&=\begin{pmatrix}
\hat{p}\hat{p}\cdots\hat{p} & 0 \\
0 & \hat{p}^\dagger\hat{p}^\dagger\cdots\hat{p}^\dagger
\end{pmatrix}.
\end{align}

Substituting this into Eq.~(\ref{Eq10}), the NLO naturally decomposes into two conjugate contributions:
\begin{equation}\label{Eq16}
\hat{S}_N = \hat{P}_N + \hat{P}_N^\dagger,
\end{equation}
where $\hat{P}_N = \hat{p} \cdot \hat{p}\hat{p} \cdots \hat{p}\hat{p} \cdot \hat{p}$ is a string-like operator. Accordingly, we employ the single-site operator $\hat{p}$ as the key object characterizing $\hat{S}_N$ across various quantum spin models.

\textit{ Adaptive Optimization Approach.}
In practice, maximizing the violation of Bell-type inequalities requires a rigorous optimization of NLO $\hat{S}_N$—or, equivalently, its constituent single-site operator $\hat{p}$. A traditional approach, as established in Refs.~\cite{Xu2024, Sun2022}, can be summarized by considering a Hamiltonian $\hat{H}(h)$ with a free parameter $h$:

(1) State Preparation: For each value of $h$, the ground state is determined in the form of a 
translation-invariant matrix product state (MPS), denoted as $|\psi_h\rangle$. 

(2) Operator Optimization: For a given $|\psi_h\rangle$, the nonlocality transfer matrix is constructed, and the optimal operator is found by solving:
\begin{equation}\label{Eq17}
\max_{\hat{p}} |\lambda_1(\hat{p}, |\psi_h\rangle)|.
\end{equation}

In this standard framework, the NLO is re-optimized for every individual state $|\psi_h\rangle$; hence, we refer to it as the Adaptive Optimization approach. While this method yields highly reliable and precise results, its primary drawback is the significant computational cost incurred by performing multiple rounds of numerical optimization across the parameter space.

\subsection{Frozen-Operator Approximation}\label{sec:approximation}

To address the efficiency constraints of the adaptive method, we propose an alternative approach termed the Frozen-Operator Approximation. This method aims to characterize the behavior of $|\lambda_1|$ as a function of $h$ with reduced numerical overhead:

(1) Reference Optimization: We first perform the full optimization as described in Eq.~\eqref{Eq17} 
at a specific reference point $h^*$, identifying the corresponding optimal operator $\hat{p}^*$.

(2) Approximate Evaluation: For a neighborhood $h\in[h^*-\delta, h^*+\delta]$ surrounding the reference point, we "freeze" the operator $\hat{p}^*$ and use it to estimate $|\lambda_1(h)|$:
\begin{equation}\label{Eq18}
\langle \psi_h | \hat{S}_N(\hat{p}^*) | \psi_h \rangle \rightarrow T_{\hat{p}^*} \rightarrow |\lambda_1(h)|.
\end{equation}

While one might expect this approximation to hold only for small values of $\delta$, our findings reveal a surprising result: for the three models investigated in this study, the approximation remains highly accurate even when $\delta$ is large enough to span the majority of the phase diagram. The Frozen-Operator Approximation thus provides high-precision results comparable to the strict Adaptive Optimization approach, yet at a drastically reduced computational expense.

\subsection{Pattern analysis on the optimal operator $\hat{p}$}\label{sec:pattern_analysis}

We will perform a detailed analysis on the optimal $\hat{p}(h)$ configurations across distinct quantum phases to isolate persistent structural features. The analysis follows a structured workflow: 

(1) With the strict Adaptive Optimization Approach, we first identify the manifold of optimal single-site operators $\hat{p}(h)$ for a given $h$, noting that the optimization often yields a non-unique solution set; 

(2) We then analyze the evolution of these sets as a function of $h$ to distill a universal, persistent structure from the model-specific patterns.

In Sec.~\ref{sec:results}, the  specific patterns of the optimal $\hat{p}$ will be reported for each  quantum model.

\section{Main Results}\label{sec:results}
Utilizing the framework established in Sec.~\ref{sec:concepts_formulas}, we systematically investigate the  optimal NLOs in the transverse-field Ising, Cluster-Ising, and extended Ising models in separate subsections below. 

\subsection{The Transverse-Field Ising model}

We begin with the transverse-field Ising model, defined by the Hamiltonian~\cite{Milman1991, Kaneyoshi1993}
\begin{align}\label{Eq19}
   \hat{H} = -\sum_{i} \sigma_{i}^{z} \sigma_{i+1}^{z}  - h \sum_{i} \sigma_{i}^{x},
\end{align}
where $h_c=1$ is the quantum critical point. The ferromagnetic (FM) phase exists for $h<1$, while the paramagnetic (PM) phase occurs for $h>1$~\cite{Campbell2013, Liu2021}. The system possesses a $\mathbb{Z}_2$ symmetry under $\pi$-rotations about the $x$-axis, which is spontaneously broken in the FM phase ($h<1$). To select a definite symmetry-broken ground state and lift the degeneracy, we introduce a small symmetry-breaking perturbation $10^{-5}\sigma_{i}^{z}$ into the Hamiltonian. Our subsequent analysis therefore distinguishes two scenarios: the symmetry-preserving case (without perturbation) and the symmetry-broken case (with perturbation).

\subsubsection{Symmetric patterns of the optimal operator $\hat{p}$}

For the symmetry-preserving Ising model with unit-cell size $u=1$, a universal structure for the optimal operator $\hat{p}(h)$ is found to be
\begin{equation}\label{Eq20}
\begin{aligned}
\boldsymbol{a}_1 =[0,\,y_1,\, z_1], \quad
\boldsymbol{a}'_1=[0,\,y_1,\,-z_1].
\end{aligned}
\end{equation}

\begin{figure}
\includegraphics[width=0.45\textwidth,keepaspectratio]{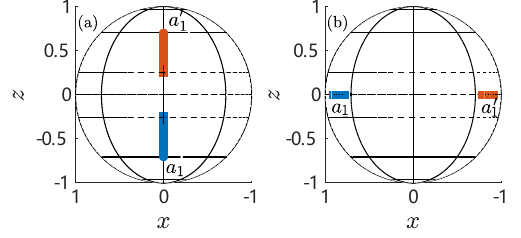}
\caption{Symmetry properties of optimal configurations for the operator $\hat{p}(\boldsymbol{a}_1, \boldsymbol{a}_1')$ in the transverse-field Ising model ($u=1$). As the transverse magnetic field $h$ varies, the optimal measurement directions $\boldsymbol{a}_1$ (blue) and $\boldsymbol{a}_1'$ (orange) trace trajectories across the spherical surface. Two distinct symmetry scenarios are investigated:  (a) Under explicit preservation of the $\mathbb{Z}_2$ symmetry in the ground states, the optimal unit vectors $\boldsymbol{a}_1$ and $\boldsymbol{a}_1'$ are confined to the $y$-$z$ plane and exhibit mirror symmetry with respect to the $y$-axis [See Eq. (\ref{Eq20})]. (b) When the $\mathbb{Z}_2$ symmetry is lifted by a weak symmetry-breaking perturbation, the optimal configurations shift to the $x$-$y$ plane while maintaining mirror symmetry about the $y$-axis [See Eq. (\ref{Eq22})].
}
\label{Fig3}
\end{figure}

The magnitudes of $y_1$ and $z_1$ depend on the external magnetic field $h$. This relation implies that the optimal unit vectors $\boldsymbol{a}_1$ and $\boldsymbol{a}'_{1}$ lie entirely in the $y$-$z$ plane and are related by mirror symmetry about the $y$-axis, with opposite $z$-components. Please see Fig.~\ref{Fig3}(a). 

For analytical tractability, we explicitly analyze the structure of the optimal operator $\hat{p}$. First, we parameterize the unit vectors in Eq.~(\ref{Eq20}) using a single angle $\theta$ in spherical coordinates: $\boldsymbol{a}_1 =[0, \cos \theta , \sin \theta ]$ and $\boldsymbol{a}'_1 =[0, \cos \theta , -\sin \theta ]$. 

Applying Eq.~(\ref{Eq12}), the optimal $\hat{p}_1$ operator for this case takes the following universal form:
\begin{align}\label{Eq21}
\hat{p}_1 = \cos{\theta}\, \sigma_y - i\sin{\theta}\, \sigma_z.
\end{align}

For the symmetry-broken Ising model, the optimal configurations instead satisfy
\begin{equation}\label{Eq22}
\begin{aligned}
\boldsymbol{a}_1 =[ x_1,\,y_1,\, 0], \quad
\boldsymbol{a}'_1=[-x_1,\,y_1,\, 0].
\end{aligned}
\end{equation}

The magnitudes of $x_1$ and $y_1$ are again determined by the field $h$. This symmetry relation implies that the optimal unit vectors are confined to the $x$-$y$ plane and exhibit mirror symmetry about the $y$-axis, with opposite $x$-components. Please see  Fig.~\ref{Fig3}(b). Consequently, it is easy to see that the optimal NLO $\hat{p}_1$ can be expressed as
\begin{align}\label{Eq23}
\hat{p}_1 = -i\cos{\theta}\, \sigma_x +\sin{\theta}\, \sigma_y.
\end{align}

The transition of the optimal measurement settings from the $y$-$z$ plane (Eq.~\ref{Eq20}) to the $x$-$y$ plane (Eq.~\ref{Eq22}) upon introducing the symmetry-breaking perturbation reveals a profound physical mechanism regarding how quantum nonlocality is maximally extracted. In the symmetry-preserving case, the ground state is a macroscopic superposition without a definite local magnetization, allowing the nonlocality operator to utilize the $z$-components to capture entanglement. However, when the $\mathbb{Z}_2$ symmetry is explicitly broken, the system acquires a definite classical local polarization (e.g., $\langle \sigma^z \rangle \neq 0$). To maximize the violation of Bell-type inequalities, the measurement axes must avoid this classical macroscopic background. Consequently, the optimal measurement plane geometrically rotates to be orthogonal to the primary axis of local polarization, shifting into the $x$-$y$ plane to exclusively probe the transverse quantum fluctuations where the genuine multipartite entanglement resides.

The optimal configurations in Eqs.~(\ref{Eq21}) and~(\ref{Eq23}) reveal that the resulting  operators $\hat{p}$ exhibit well-defined structured patterns, regardless of whether the $\mathbb{Z}_2$ symmetry is preserved or broken.

\subsubsection{Robustness of the optimal operator $\hat{p}$}
In this subsection, we evaluate the robustness of the optimal operator $\hat{p}$ against variations in the magnetic field $h$ by comparing the nonlocality spectra obtained via the Adaptive Optimization and Frozen-Operator Approximation approaches. 

\begin{figure}
\includegraphics[width=0.45\textwidth,keepaspectratio]{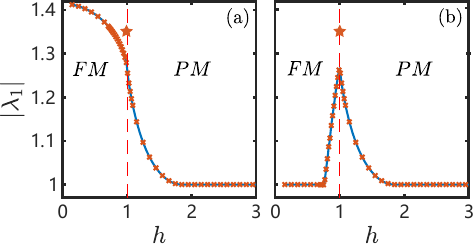}
\caption{Robustness of the optimal operator $\hat{p}$ in the transverse-field Ising chains. The red dashed lines denote the phase boundaries. We compare the numerical results of the leading eigenvalue $|\lambda_1|$ of the nonlocality transfer matrix by the Adaptive Optimization approach (blue solid lines), where the operator $\hat{p}$ is optimized independently for every magnetic field $h$, with the Frozen-Operator Approximation (orange crosses). In the latter, the operator is optimized solely at the reference point $h^*=1$ (indicated by the star), with that single configuration then applied across the full range of $h$. Panels (a) and (b) display results for the $\mathbb{Z}_2$ symmetry-preserved and $\mathbb{Z}_2$ symmetry-broken situations, respectively. The near-perfect overlap between the two approaches reveals a striking result: the optimal operator configuration is remarkably robust across the entire parameter space.}
\label{Fig4}
\end{figure}

Our numerical results are presented in Figure~\ref{Fig4}. The blue solid lines represent $|\lambda_1|$ calculated using the conventional Adaptive Optimization approach. This method is computationally demanding, as it requires independent multi-variable optimizations for each of the $90$ sampled field values, with $10$ repeated trials per point to guarantee numerical stability.

In contrast, the orange crosses denote results obtained via the Frozen-Operator Approximation. Under this scheme, the single-site operator $\hat{p}$ is optimized only once at the critical reference point $h^*=1$ (indicated by the star); this fixed configuration is then applied across the entire range of field values. 

Remarkably, the results from both methods exhibit near-perfect overlap across the entire phase diagram, spanning both the FM and PM phases. This agreement is not confined to the neighborhood of the reference point $h^*=1$ but persists throughout the parameter space. Notably, as shown in Fig.~\ref{Fig4}(a) and~\ref{Fig4}(b), this behavior remains consistent regardless of whether $\mathbb{Z}_2$ symmetry is preserved. This overlap underscores the inherent robustness of the optimal operator $\hat{p}$, demonstrating that a single configuration determined at the critical point is sufficient to achieve the maximum $|\lambda_1|$ across the model's entire parameter space.

This robustness warrants further discussion. First, the phase-independent universal patterns of the optimal $\hat{p}$ in Eqs.~(\ref{Eq20}) and~(\ref{Eq22}) already identify an underlying structural stability of the optimal nonlocality operators.

Building on this, the overlap shown in Fig.~\ref{Fig4}  reveals that the specific configuration of the operator $\hat{p}$, rather than the precise value of the parameter $\theta$, plays the central role in achieving the maximum violation of Bell-type inequalities.

\subsection{The Cluster-Ising Model}

We now investigate the Cluster-Ising model, a paradigmatic system hosting a rich interplay of symmetry-protected topological order and conventional long-range order. Its Hamiltonian reads~\cite{Wolf2006, Pachos2004, Ding2019, Giampaolo2014}
\begin{align}\label{Eq24}
   \hat{H} = -\sum_{i} \left(\sigma_{i - 1}^{x} \sigma_{i}^{z} \sigma_{i + 1}^{x} + J\sigma_{i}^{x} \sigma_{i + 1}^{x} + h\sigma_{i}^{z}\right),
\end{align}
where $J \ge 0$ is the strength of the nearest-neighbor Ising interaction and $h$ is the external transverse magnetic field. The three terms drive the system toward distinct phases: the three-body cluster term stabilizes a ground state with non-trivial topological order (the cluster phase); the Ising term favors ferromagnetic ordering; and the field term favors a trivial polarized state. Accordingly, the phase diagram comprises three regions: a topologically ordered cluster (CL) phase, a $\mathbb{Z}_2$ ferromagnetic Ising FM phase, and a trivial paramagnetic (PM) phase. For fixed $J$, the CL phase occurs for $h < 1-J$, the Ising FM phase for $1-J < h < 1+J$, and the PM phase for $h > 1+J$~\cite{Skroevseth2009, Son2011}.

For this model, we also consider two scenarios: the symmetry-preserving case and the symmetry-broken case, with the latter induced by the weak perturbation $10^{-5}\sigma_{i}^{z}$. This perturbation is sufficiently small to leave the phase boundaries unaffected.

\subsubsection{Symmetric patterns of the optimal operator $\hat{p}$}

\begin{figure}
\includegraphics[width=0.45\textwidth,keepaspectratio]{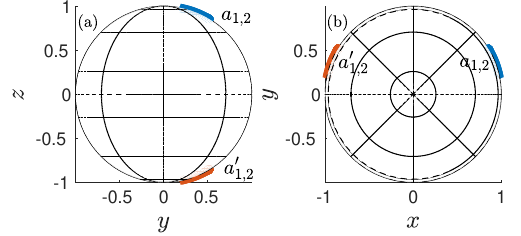}
\caption{Symmetry properties of optimal configurations for the operators $\hat{p}_1(\boldsymbol{a}_1, \boldsymbol{a}_1')$ and $\hat{p}_2(\boldsymbol{a}_2, \boldsymbol{a}_2')$ in the Cluster-Ising model with $J=0.3$ ($u=2$). As the transverse magnetic field $h$ varies, the optimal measurement directions $\boldsymbol{a}_{1,2}$ (blue, with $\boldsymbol{a}_1=\boldsymbol{a}_2$) and $\boldsymbol{a}_{1,2}'$ (orange, with $\boldsymbol{a}_1'=\boldsymbol{a}_2'$) trace trajectories across the spherical surface. Two distinct symmetry scenarios are investigated:  (a) Under explicit preservation of the $\mathbb{Z}_2$ symmetry in the ground states, the optimal unit vectors $\boldsymbol{a}_{1,2}$ and $\boldsymbol{a}_{1,2}'$ are confined to the $y$-$z$ plane and exhibit mirror symmetry with respect to the $y$-axis [See Eq. (\ref{Eq25})]. (b) When the $\mathbb{Z}_2$ symmetry is lifted by a weak symmetry-breaking perturbation, the optimal configurations shift to the $x$-$y$ plane while maintaining mirror symmetry about the $y$-axis [See Eq. (\ref{Eq28})]. Qualitatively similar symmetry behavior is observed for other coupling strengths, such as $J=0.5$ and $J=1$.        
}
\label{Fig5}
\end{figure}

In the symmetry-preserving Cluster-Ising model ($J>0$) with a unit-cell size of $u=2$, the optimal operator $\hat{p}(h)$ exhibits a universal structure:
\begin{equation}\label{Eq25}
\begin{aligned}
\boldsymbol{a}_1 =[ 0,\,y_1,\, z_1], \quad
\boldsymbol{a}'_1=[0,\,y_1,\, -z_1];\\
\boldsymbol{a}_2 =[ 0,\,y_1,\, z_1], \quad
\boldsymbol{a}'_2=[0,\,y_1,\, -z_1].
\end{aligned}
\end{equation}

The magnitudes of $y_1$ and $z_1$ depend on the external magnetic field $h$. We observed that $\boldsymbol{a}_2$ ($\boldsymbol{a}'_2$) is identical to $\boldsymbol{a}_1$ ($\boldsymbol{a}'_1$). This identity indicates that both sublattice sites share an identical, mirror-symmetric optimal measurement structure. The optimal unit vectors $(\boldsymbol{a}_i, \boldsymbol{a}'_i)$ for both sublattice sites lie in the $y$-$z$ plane and are related by mirror symmetry about the $y$-axis, with their $z$-components being equal and opposite, as visualized in Fig.~\ref{Fig5}(a). The corresponding optimal operators take the form as
\begin{equation}\label{Eq26}
\hat{p}_1 = \hat{p}_2=\cos{\theta}\,\sigma_y - i\sin{\theta}\,\sigma_z.
\end{equation}

In the special case $J=0$, the second sublattice decouples, and the unit cell effectively reduces to $u=1$ with $\boldsymbol{a}_2 =\boldsymbol{a}'_2=[ 0,0, 1]$. The optimal operators then become
\begin{equation}\label{Eq27}
\begin{aligned}
\hat{p}_1& =\cos{\theta}\,\sigma_y - i\sin{\theta}\,\sigma_z, \\
\hat{p}_2&=\sigma_z.
\end{aligned}
\end{equation}

This transition from the uniform $u=2$ structure [Eq.~(\ref{Eq26})] to the decoupled $u=1+1$ structure [Eq.~(\ref{Eq27})] as $J\to 0$ is continuous (up to a global phase factor), reflecting the smooth vanishing of the Ising interaction between sublattices.

When the $\mathbb{Z}_2$ symmetry is broken by a perturbation, the universal structure of the optimal operators $\hat{p}(h)$ instead satisfy
\begin{equation}\label{Eq28}
\begin{aligned}
\boldsymbol{a}_1 =[ x_1,\,y_1,\, 0], \quad
\boldsymbol{a}'_1=[ -x_1,\,y_1,\, 0];\\
\boldsymbol{a}_2 =[x_1,\,y_1,\, 0], \quad
\boldsymbol{a}'_2=[-x_1,\,y_1,\, 0].
\end{aligned}
\end{equation}

The magnitudes of $x_1$ and $y_1$ depend on the external magnetic field $h$. Notably, the symmetry breaking rotates the optimal measurement plane from $y$-$z$ to $x$-$y$. Both sublattice vector pairs remain confined to this new plane and maintain mirror symmetry about the $y$-axis, with their $x$-components being equal and opposite, As depicted in Fig.~\ref{Fig5}(b). Consistent with the physical picture discussed in the transverse-field Ising model, this geometric rotation reflects the operators' optimization strategy to capture transverse quantum fluctuations and bypass the explicit local magnetization induced by the symmetry-breaking perturbation. The corresponding optimal operators $\hat{p}$ take the universal form 
\begin{equation}\label{Eq29}
\hat{p}_1=\hat{p}_2=-i\cos{\theta}\,\sigma_x+\sin{\theta}\,\sigma_y.
\end{equation}

For the Cluster-Ising model with three-site interactions and unit cell size $u$ exceeds 1, Eqs. (\ref{Eq26}) and (\ref{Eq29}) reveals a compelling universal conclusion: the optimal NLOs $\hat{p}$ consistently maintain a structured, mirror-symmetric configuration, even when subjected to symmetry-breaking perturbations.

\begin{figure}
\includegraphics[width=0.45\textwidth,keepaspectratio]{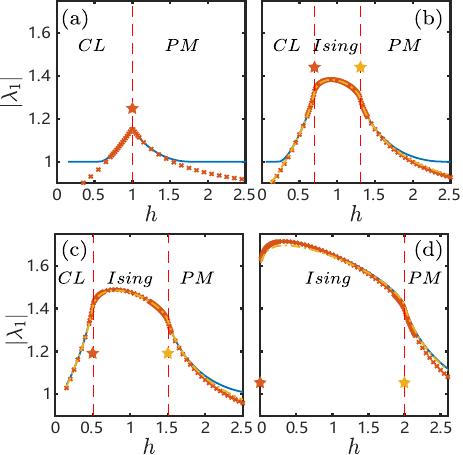}
\caption{Robustness of the optimal operator $\hat{p}$ in  the Cluster-Ising model by considering  $\mathbb{Z}_2$ symmetry-preserving ground states. The red dashed lines denote the phase boundaries. We compare the leading eigenvalue $|\lambda_1|$ by the Adaptive Optimization approach (blue solid curves) with the Frozen-Operator Approximation: orange crosses use configurations optimized at the reference point $h^* = 1-J$ (orange star), while yellow dashed curves use those from the reference point $h^* = 1+J$ (yellow star). The high degree of overlap across most regions in the parameter space demonstrates that the optimal operator configuration is remarkably robust.
}
\label{Fig6}
\end{figure}

\subsubsection{Robustness of the optimal operator $\hat{p}$}

In this subsection, we continue the comparison between the conventional Adaptive Optimization and the Frozen-Operator Approximation, probing the robustness of the optimal operator $\hat{p}$ against variations in the magnetic field $h$.

\begin{figure}
\includegraphics[width=0.45\textwidth,keepaspectratio]{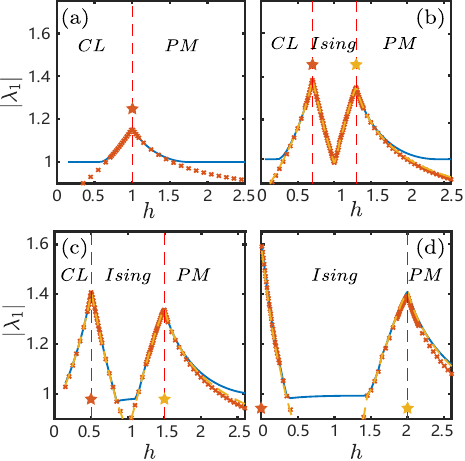}
\caption{Robustness of the optimal operator $\hat{p}$ in  the Cluster-Ising model by considering  $\mathbb{Z}_2$ symmetry-broken ground states. The red dashed lines denote the phase boundaries. The blue solid curves denote the {Adaptive Optimization} approach, while orange crosses and yellow dashed curves represent the {Frozen-Operator Approximation} optimized at $h^* = 1-J$ and $h^* = 1+J$ (stars), respectively. All other parameter settings are identical to those in Fig.~\ref{Fig6}. The consistent overlap in most regions of the phase space underscores the robustness of the optimal operator configuration.
}
\label{Fig7}
\end{figure}

Figure~\ref{Fig6} illustrates the evolution of $|\lambda_1|$ with the external field $h$ for several coupling strengths $J$ in the $\mathbb{Z}_2$ symmetry-preserving Cluster-Ising model.  Fig.~\ref{Fig7} displays the effects of symmetry breaking. In both Figs.~\ref{Fig6} and~\ref{Fig7}, the blue solid lines represent the Adaptive Optimization results. This approach requires at least $10$ independent trials per data point across $100$ or more sample values, incurring substantial computational overhead.

In contrast, our efficient alternative approach selects the optimal operators $\hat{p}$ at the critical points $h^*=1 \pm J$ (marked by star) and applies these fixed configurations across the entire field range. For the case of $J=0$, the optimal $\hat{p}$ extracted at $h^*=1$ is applied uniformly [orange crosses, Fig.~\ref{Fig6}(a) and Fig.~\ref{Fig7}(a)]. For $J>0$, the optimal $\hat{p}$ obtained at $h^*=1\pm J$ generate the results shown as orange crosses and yellow dashed lines in panels (b)–(d) of Figs.~\ref{Fig6} and ~\ref{Fig7}. Evidently, these results from both method exhibit near-perfect overlap across most of the phase diagram, spanning the transition from the CL phase to the PM phases.  Such a close agreement demonstrates that a single set of optimal operators, $\hat{p}$, is sufficient to accurately identify the global maximum of $|\lambda_1|$ throughout the majority of the phase space.

These coincident findings further confirm that the specific structural configuration of the operator $\hat{p}$, rather than the precise numerical tuning of the parameter $\theta$, is the primary driver in achieving the maximum violation of Bell-type inequalities. Furthermore, this "frozen" approach offers a significant reduction in computational cost without sacrificing physical accuracy, highlighting the underlying stability of the optimal nonlocality operators.

\subsection{The Extended Ising Model}
Now, we consider the extended quantum Ising model with unit-cell size $u=2$, described by the Hamiltonian~\cite{Zhang2015, Zhang2018}
\begin{align}\label{Eq30}
\hat{H}&=\sum_{i }\left(\frac{1 + \gamma}{2}\sigma_i^x\sigma_{i + 1}^x+\frac{1 - \gamma}{2}\sigma_i^y\sigma_{i + 1}^y\right)\nonumber\\
&+\alpha\sum_{i}\sigma_i^z\left(\frac{1 + \delta}{2}\sigma_{i - 1}^x\sigma_{i + 1}^x+\frac{1 - \delta}{2}\sigma_{i - 1}^y\sigma_{i + 1}^y\right)\nonumber\\
&+h\sum_{i}\sigma_i^z.
\end{align}

This Hamiltonian encompasses interactions governed by four key parameters: (i) the transverse magnetic field $h$; (ii) the three-spin coupling strength $\alpha$; (iii) the next-nearest-neighbor anisotropy parameter $\delta$; and (iv) the nearest-neighbor anisotropy parameter $\gamma$. The extended Ising model possesses a rich phase diagram with multiple topological phases, reflecting the interplay among these four parameters. We analyze each in turn.

\begin{figure*}
\includegraphics[width=0.9\textwidth,keepaspectratio]{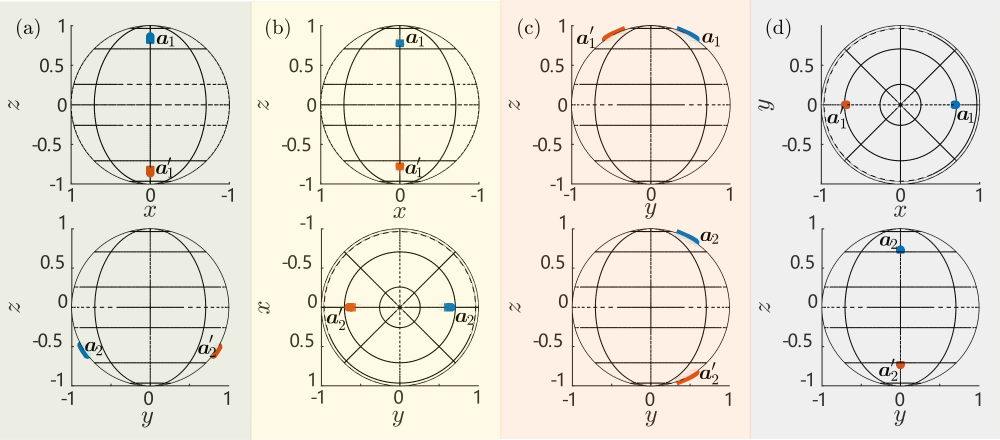}
\caption{Symmetry properties of optimal configurations for the operators $\hat{p}_1(\boldsymbol{a}_1, \boldsymbol{a}_1')$ and $\hat{p}_2(\boldsymbol{a}_2, \boldsymbol{a}_2')$ in the extended Ising model ($u=2$). The Hamiltonian is characterized by four parameters: the transverse magnetic field $h$, the three-spin coupling $\alpha$, the anisotropy $\delta$, and the nearest-neighbor anisotropy $\gamma$. Four distinct parameter sweeps are presented: (a) varying $h$ while fixing $\{\alpha, \delta, \gamma\}$; (b) varying $\alpha$ while fixing $\{h, \delta, \gamma\}$; (c) varying $\delta$ while fixing $\{h, \alpha, \gamma\}$; and (d) varying $\gamma$ while fixing $\{h, \alpha, \delta\}$. In each case, the upper panel displays the trajectories of the optimal unit vectors $\boldsymbol{a}_1$ and $\boldsymbol{a}'_1$, while the lower panel displays those for $\boldsymbol{a}_2$ and $\boldsymbol{a}'_2$. Across all investigated scenarios, the optimal configurations consistently exhibit mirror symmetry between the vector pairs $\boldsymbol{a}_i$ and $\boldsymbol{a}'_i$.
}
\label{Fig8}
\end{figure*}

\subsubsection{Symmetric patterns of the optimal operator $\hat{p}$}

(i) \textit{Transverse magnetic field $h$}: We fix $\gamma=1$, $\delta=1$, and $\alpha=1.5$. The model undergoes topological quantum phase transitions at $h_{c1}=0.5$ and $h_{c2}=2.5$~\cite{Yin2019}. The optimal configurations universally satisfy
\begin{equation}\label{Eq31}
\begin{aligned}
\boldsymbol{a}_1 =[ 0,\,y_1,\, z_1], \quad
\boldsymbol{a}'_1=[ 0,\,y_1,\, -z_1];\\
\boldsymbol{a}_2 =[ 0,\,y_1,\, z_1], \quad
\boldsymbol{a}'_2=[ 0,\,-y_1,\, z_1].
\end{aligned}
\end{equation}

Here, $\boldsymbol{a}_1$ and $\boldsymbol{a}'_1$ lie in the $y$-$z$ plane with mirror symmetry about the $y$-axis, while $\boldsymbol{a}_2$ and $\boldsymbol{a}'_2$ also lie in the $y$-$z$ plane but are symmetric about the $z$-axis, as seen in Fig.~\ref{Fig8}(a). The corresponding optimal operators are
\begin{equation}\label{Eq32}
\begin{aligned}
\hat{p}_1 &= \cos{\theta}\, \sigma_y - i\sin{\theta}\, \sigma_z,\\
\hat{p}_2 &= -i\cos{\theta}\, \sigma_y +\sin{\theta}\, \sigma_z.
\end{aligned}
\end{equation}

(ii) \textit{Three-spin coupling $\alpha$}: We fix $\gamma=1$, $\delta=-1$, and $h=1$. Topological phase transitions occur at $\alpha_{c1}=(-\sqrt{5}-1)/2$, $\alpha_{c2}=0$, $\alpha_{c3}=(\sqrt{5}-1)/2$, and $\alpha_{c4}=2$~\cite{Yin2019}. As visualized in Fig.~\ref{Fig8}(b), the optimal NLOs satisfy the same structure as in case (i):
\begin{equation}\label{Eq33}
\begin{aligned}
\boldsymbol{a}_1 =[ 0,\,y_1,\, z_1], \quad
\boldsymbol{a}'_1=[ 0,\,y_1,\, -z_1];\\
\boldsymbol{a}_2 =[ 0,\,y_1,\, z_1], \quad
\boldsymbol{a}'_2=[ 0,\,-y_1,\, z_1],
\end{aligned}
\end{equation}
with $y_1$ and $z_1$ now determined by $\alpha$ rather than $h$. The optimal operators are
\begin{equation}\label{Eq34}
\begin{aligned}
\hat{p}_1 &= \cos{\theta}\, \sigma_y - i\sin{\theta}\, \sigma_z,\\
\hat{p}_2 &= -i\cos{\theta}\, \sigma_y +\sin{\theta}\, \sigma_z.
\end{aligned}
\end{equation}

(iii) \textit{Next-nearest-neighbor anisotropy $\delta$}: We fix $\gamma=1$, $\alpha=1$, and $h=-0.3$. The symmetry relationship can be described by Fig.~\ref{Fig8}(c). Topological phase transitions occur at $\delta_{c1}=-1.2747$ and $\delta_{c2}=0.5604$~\cite{Yin2019}. The optimal configurations are
\begin{equation}\label{Eq35}
\begin{aligned}
&\boldsymbol{a}_1 =[ 0,\,y_1,\, z_1], \quad
\boldsymbol{a}'_1=[ 0,\,-y_1,\, z_1];\\
&\boldsymbol{a}_2 =[ 0,\,-y_1,\, z_1], \quad
\boldsymbol{a}'_2=[ 0,\,-y_1,\, -z_1].
\end{aligned}
\end{equation}

The corresponding optimal operators are
\begin{equation}\label{Eq36}
\begin{aligned}
\hat{p}_1 &= \cos{\theta}\, \sigma_y - i\sin{\theta}\, \sigma_z,\\
\hat{p}_2 &= -\cos{\theta}\, \sigma_y -i\sin{\theta}\, \sigma_z.
\end{aligned}
\end{equation}

(iv) \textit{Nearest-neighbor anisotropy $\gamma$}: We fix $\delta=1$, $h = -0.5$, and $\alpha = 1$. Figure~\ref{Fig8}(d) provides a graphical representation of the symmetry relationship. Topological phase transitions occur at $\gamma_{c1}=-0.618$ and $\gamma_{c2}=1.618$~\cite{Li2018}. The optimal configurations satisfy
\begin{equation}\label{Eq37}
\begin{aligned}
\boldsymbol{a}_1 =[ x_1,\,0,\, z_1], \quad
\boldsymbol{a}'_1=[ -x_1,\,0,\, z_1];\\
\boldsymbol{a}_2 =[ x_1,\,0,\, z_1], \quad
\boldsymbol{a}'_2=[ x_1,\,0,\, -z_1].
\end{aligned}
\end{equation}

The corresponding optimal operators are
\begin{equation}\label{Eq38}
\begin{aligned}
\hat{p}_1 &= -i\cos{\theta}\, \sigma_x +\sin{\theta}\, \sigma_z,\\
\hat{p}_2 &= \cos{\theta}\, \sigma_x -i\sin{\theta}\, \sigma_z.
\end{aligned}
\end{equation}

Across all four parameter scans of the extended Ising model, the optimal NLO $\hat{p}$ consistently assumes a well-structured, mirror-symmetric form, confirming the universality of the underlying rules governing optimal measurement configurations.

\begin{figure}
\includegraphics[width=0.45\textwidth,keepaspectratio]{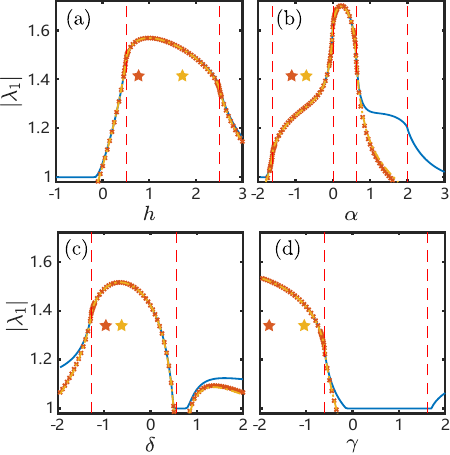}
\caption{Robustness of the optimal operator $\hat{p}$ in the extended Ising model. The leading eigenvalue $|\lambda_1|$ is shown as a function of (a) $h$, (b) $\alpha$, (c) $\delta$, and (d) $\gamma$, with all other parameters held constant. The phase boundaries are indicated by red dashed lines. Blue solid curves represent the leading eigenvalue $|\lambda_1|$  by the Adaptive Optimization approach, and orange crosses and yellow dashed curves denote results by the Frozen-Operator Approximation. The frozen configurations were optimized at some random points in the quantum phase marked by stars. The high degree of overlap across extensive parameter regions  demonstrates that the optimal operator configuration is remarkably robust.
}
\label{Fig9}
\end{figure}

\subsubsection{Robustness of the optimal operator $\hat{p}$}
We further extend our analysis of operator robustness to the extended Ising model, which features a more complex parameter space. Figure~\ref{Fig9} illustrates the evolution of $|\lambda_1|$ as a function of the Hamiltonian parameters: (a) the transverse field $h$, (b) the three-body interaction $\alpha$, (c) the next-nearest-neighbor interaction $\delta$, and (d) the nearest-neighbor interaction$\gamma$.

To determine the leading eigenvalue $|\lambda_1|$ of the nonlocality transfer matrix, we again contrast two strategies. The Adaptive Optimization approach (solid blue lines) serves as our benchmark but requires a significant computational investment—necessitating at least 10 independent trials per data point across 100 or more sampled values to ensure convergence.

To rigorously test the robustness of the optimal operator $\hat{p}$, we apply the Frozen-Operator Approximation under even more stochastic conditions. In this instance, rather than selecting $\hat{p}$ at a critical point, we fix the operator at an arbitrary point within a specific phase (indicated by "star" symbols). Despite this arbitrary selection, the resulting curves (markers) demonstrate excellent agreement with the Adaptive Optimization results across most regions.

The recurrence of this phenomenon confirms that the consistent performance of "frozen" operators is not unique to simple models. Instead, it highlights the intrinsic physical significance and structural universality of the optimal operator $\hat{p}$, suggesting that its underlying configuration remains fundamentally stable across the diverse phases of the extended Ising model.

\section{Summary and Discussions}\label{sec:conclusion}

Previous research on multipartite nonlocality in the ground states of one-dimensional quantum chains has primarily focused on the nonlocality measure $S$ and its spectrum $\vert\lambda_i\vert$. However, the underlying optimal NLO $\hat{S}_N$ has remained largely elusive. In this work, we have explicitly investigated the configurations of the string-like optimal NLOs, characterized by the optimal single-site operator $\hat{p}$. By analyzing the infinite-size transverse-field Ising model, the Cluster-Ising model, and the extended Ising model, we have established two principal findings.

First, the optimal configurations exhibit an intrinsic mirror symmetry: each pair measurement directions $\{\boldsymbol{a}_i, \boldsymbol{a}'_i\}$ within $\hat{p}$ is related by reflection about a principal axis. Consequently, the optimal $\hat{p}$  operator adopts a remarkably compact form. For instance, in the $\mathbb{Z}_2$-symmetric ground states of the transverse-field Ising model, we find $\hat{p}_1 = \cos{\theta}\, \sigma_y - i\sin{\theta}\, \sigma_z$.

Similar structures are observed across all three models and persist even under explicit symmetry-breaking perturbations. This suggests that mirror symmetry is an intrinsic property for the optimal NLO, rather than an artifact of the system's Hamiltonian symmetry. This finding renders the optimal nonlocality operator both physically transparent and experimentally more accessible.

Second, the optimal $\hat{p}$ operator is remarkably robust throughout the phase diagram of the models. A single fixed $\hat{p}$, optimized at a single reference point (such as the critical point), suffices to reproduce the global maximum of $|\lambda_1|$ across most regions in the  phase diagram. This robustness leads to several significant implications:

(1) Configuration over Fine-Tuning: From a physical perspective, this robustness indicates that the specific configuration of the operator $\hat{p}$, rather than the fine-tuning of its parameter (i.e., $\theta$), is the primary driver for achieving the maximum violation of Bell-type inequalities. 

(2) A New Optimization Paradigm: It will reshape the numerical optimization paradigm for multipartite nonlocality. While the conventional Adaptive Optimization approach is computationally demanding, the demonstrated robustness of the optimal $\hat{p}$ operator justifies the Frozen-Operator Approximation. This approach provides high-precision results comparable to strict optimization at a drastically reduced computational expense, facilitating the study of more complex systems.

(3) Origin of Singularities: This robustness clarifies the origin of singularities in the nonlocality measure $S(h)=\max \langle \psi_h | \hat{S}_N \vert \psi_h \rangle$  at quantum phase transitions. It was previously unclear whether such singularities arose from dramatic changes in the ground state $|\psi_h\rangle$, or both $|\psi_h\rangle$ and  the adaptively optimized $\hat{S}_N$. Our finding reveals that the singularity in $S(h)$ stems purely from the evolution of the ground state $|\psi_h\rangle$, as the optimal operator is robust across the transitions between distinct quantum phases.

(4) Insight into Topological Phases: Furthermore, the remarkable robustness of the optimal $\hat{p}$ configuration in the Cluster-Ising and extended Ising models provides a deep physical insight into topological quantum phases. In these symmetry-protected topological phases, the ground states are characterized by non-local string order parameters rather than local magnetizations. The optimal string-like NLO $\hat{P}_N$ essentially adapts to and "learns" this underlying topological string order~\cite{Verresen2018, Yu2022}. The geometric stability of $\hat{p}$ across the phase diagram implies that the fundamental structure of genuine multipartite entanglement is topologically protected, remaining robust against local Hamiltonian perturbations as long as the relevant symmetry is not violently destroyed.

(5) Simplifying Experimental Implementation: Measuring global string operators in large-scale quantum simulators, such as Rydberg atom arrays~\cite{Semeghini2021, Ebadi2021} or superconducting qubits~\cite{Kim2023, Satzinger2021}, typically suffers from extensive calibration and sampling overhead, particularly if the local measurement axes must be adaptively tuned for different quantum states. Our finding that the optimal $\hat{p}$ is highly structured and parameter-independent provides a crucial advantage: it allows experimentalists to utilize a single, pre-determined set of local measurement bases across different quantum phases. By restricting the measurements to these specific robust configurations, the experimental complexity associated with multi-variable optimization is substantially alleviated. This insight paves the way for integrating our frozen-operator approach with modern efficient measurement protocols, such as shadow tomography or randomized measurements~\cite{McGinley2022, Elben2023, Huang2020a, Bluvstein2022, Hu2025, Wu2026, King2025, Zhang2025}, enabling the detection of macroscopic Bell violations in near-term quantum devices.

In conclusion, these findings provide a fundamental understanding of the optimal nonlocality operators in quantum chains. They redefine the numerical optimization paradigm for multipartite nonlocality, and provide a robust and experimentally actionable framework for designing high-fidelity Bell tests in practical laboratory protocols.

While the Frozen-Operator Approximation demonstrates remarkable success across the phase diagrams studied here, we note that it inherently assumes a degree of continuity in the system's correlation structure. Transitions into radically different universality classes or highly excited states, where the entanglement geometry undergoes an abrupt global reconstruction, may still necessitate a re-optimization of the reference point. Several natural extensions warrant future investigation. The robustness of $\hat{p}$ identified here, geometric stability under Hamiltonian parameter variation, is conceptually complementary to the fragility of optimal settings reported in Ref.~\cite{Aloy2026} for spin-1 Bell inequalities, where the measure of integrable configurations vanishes with system size. Clarifying whether the mirror-symmetry structure underlies analogous spectral phenomena in spin-chain Bell operators would bridge these two perspectives. Furthermore, extending the present framework to systems with larger unit cells, more complex lattice geometries (e.g., 2D tensor networks), non-Hermitian settings, or finite-temperature mixed states constitutes a promising research agenda.

\section*{Acknowledgments}

This work is supported by the National Natural Science Foundation of China (11975175) and the Fundamental Research Funds for the Central Universities (104972025KFYjc0079).

\section*{Data Availability}

All raw data corresponding to the findings reported in this paper are available from the corresponding author upon reasonable request.

\section*{Author declarations}

The authors have no conflicts of interest to disclose.

\bibliography{references_main.bib}

\end{document}